\documentclass[letterpaper]{article}
\usepackage{aaai}
\usepackage{times}
\usepackage{helvet}
\usepackage{courier}
\usepackage{epsfig}
\usepackage{amssymb}
\title{Lack of Finite Characterizations for the Distance-based Revision\thanks{Thanks to Karl Schlechta for very valuable discussions.}}
\author{Jonathan Ben-Naim
\\
LIF, CNRS
\\
CMI 39, rue Joliot-Curie,
\\
F-13453 Marseille Cedex 13, France
\\
jbennaim@lif.univ-mrs.fr}

\newtheorem{definition}{Definition}
\newtheorem{notation}[definition]{Notation}
\newtheorem{proposition}[definition]{Proposition}
\makeatletter
\renewcommand{\@begintheorem}[2]{ 
\trivlist\item[\hskip\labelsep{\bf #1\ #2}]}
\renewcommand{\@opargbegintheorem}[3]{\trivlist
\item[\hskip \labelsep{\bf #1\ #2\ (#3)}]}
\makeatother
\newtheorem{proof}{Proof}

\newcommand{\qed}{\nobreak \ifvmode \relax \else
\ifdim\lastskip<1.5em \hskip-\lastskip
\hskip1.5em plus0em minus0.5em \fi \nobreak
\vrule height0.75em width0.5em depth0.25em\fi}

\newcommand{\M}[1]{M_{#1}}

\newcommand{\T}[1]{T(#1)}

\newcommand{\C}[2]{{#1}(#2)}

\begin{document}
\maketitle

\begin{abstract}
Lehmann, Magidor, and Schlechta
developed an approach to belief revision based on distances between any two valuations.
Suppose we are given such a distance $\cal D$.
This defines an operator $|_{\cal D}$, called a {\it distance operator}, which transforms any two sets of valuations $V$ and $W$
into the set $V |_{\cal D} W$ of all those elements of $W$ that are closest
to $V$.
This operator $|_{\cal D}$
defines naturally the revision of $K$ by $\alpha$ as the set of all
formulas satisfied in $\M{K} |_{\cal D} \M{\alpha}$ (i.e. the set of all those models of
$\alpha$ that are closest to the models of $K$).
This constitutes a {\it distance-based revision operator}.
Lehmann {\it et al.} characterized families
of them using a ``loop'' condition of arbitrarily big size.
An interesting question is whether this loop condition can be replaced by a finite one.
Extending the results of Schlechta, we will provide elements of negative answer.
In fact, we will show that for families of distance operators,
there is no ``normal'' characterization.
Approximatively, a characterization is normal iff it contains only
finite and universally quantified conditions.
Though they are negative, these results have an interest of their own for they
help to understand more clearly the limits of what is possible in this area.
In addition, we are quite confident that they can be used to show that for families of distance-based revision operators, there is no normal characterization either.
For instance, the families of Lehmann {\it et al.} might well be concerned with this, which suggests that their
big loop condition cannot be replaced by a finite and universally quantified condition.
\end{abstract}

\section{Introduction}

Belief revision is the study of how an intelligent agent may replace its current
epistemic state by another one which is non-trivial and incorporates new information.
In \cite{AlchourronGardenforsMakinson1}, the well-known AGM approach was proposed.
An epistemic state is modelled there by a deductively closed set of formulas $K$
and new information by a formula $\alpha$.
A revision operator is then a function that transforms $K$ and $\alpha$ into
a new set of formulas (intuitively, the revised epistemic state).

One of the contributions of the AGM approach is that
it provides well-known postulates that any reasonable revision operator should satisfy.
These postulates have been defended by their authors.
But, doubts have been expressed as to their ``soundness'', e.g.~\cite{KatsunoMendelzon1}, and especially ``completeness'', e.g.~\cite{FreundLehmann1}, \cite{DarwichePearl1}, \cite{Lehmann3}, and \cite{DarwichePearl2}.
In particular, to be accepted, an operator never needs to put some coherence
between the revisions of two different sets $K$ and $K'$.
As a consequence, some operators are accepted though they are not well-behaved when iterated.
In addition, modelling an epistemic state by just a deductively closed set of formulas
has been rejected by many researchers, e.g. \cite{BoutilierGoldszmidt1}, \cite{Boutilier1}, \cite{DarwichePearl2}, \cite{Williams1},
and \cite{NayakFooPagnuccoSattar1}. In \cite{Lehmann3} and \cite{FriedmanHalpern1},
it is argued that this modelling is not sufficient in many AI applications.

This provides motivations for another approach,
based on distances between any two valuations, introduced in \cite{LehmannMagidorSchlechta2} and
investigated further in \cite{LehmannMagidorSchlechta1}.
Their approach is in line with the AGM modelling of an epistemic state,
but it defines well-behaved iterated revisions. More precisely,
suppose we have at our disposal a distance $\cal D$ between any two valuations.
This defines an operator $|_{\cal D}$, called a {\it distance operator}, which transforms any ordered pair $(V,W)$
of sets of valuations into the set $V |_{\cal D} W$ of all those elements of $W$ that are closest
to $V$ according to ${\cal D}$.

This operator $|_{\cal D}$
defines naturally the revision of $K$ by $\alpha$ as the set of all
formulas satisfied in $\M{K} |_{\cal D} \M{\alpha}$ (i.e. the set of all those models of
$\alpha$ that are closest to the models of $K$).
This constitutes a {\it distance-based revision operator}, which is interesting for its natural aspect
and for it is well-behaved when iterated.
This is due to the fact that the revisions of the different $K$'s are all defined by the same distance,
which ensures a strong coherence between them.
Note that this is not the case with other definitions.
For instance, with sphere systems \cite{Grove1} and epistemic entrenchment relations
\cite{GardenforsMakinson1}, the revision of  each $K$ is defined by
a different structure without any ``glue'' relating them.

In \cite{LehmannMagidorSchlechta1}, several families
of distance-based revision operators were characterized by the AGM postulates together with
new ones that deal with iterated revisions. However, the latter postulates include a ``loop'' condition of
arbitrarily big size. An interesting question is whether it can be replaced by a finite condition.
Elements of negative answer were provided in \cite{Schlechta5}.
Approximatively, Schlechta call normal a characterization containing only conditions which are
finite, universally quantified (like e.g. the AGM postulates),
and simple (i.e. using only elementary operations like e.g. $\cup$, $\cap$, $\setminus$).
Then, he showed that for families of distance operators,
there is no normal characterization.

Now, there is a strong connexion between the distance operators
(which apply to valuations) and the distance-based revision operators (which apply to formulas).
It is quite reasonable to think that the work of Schlechta can be continued
to show that for families of distance-based revision operators, there is no
normal characterization either. The families investigated in \cite{LehmannMagidorSchlechta1}
might well be concerned with this, which suggests that the arbitrarily big loop condition
cannot be replaced by a finite, universally quantified, and simple condition.

The contribution of the present paper is to extend the work of Schlechta in two directions.
First, we will use the word ``normal'' in a larger sense.
Indeed, we will call normal a characterization containing only conditions which are
finite and universally quantified, but not necessarily simple
(i.e. the conditions can involve complex structures or functions, etc., we are not limited to elementary operations).
Then, we will show that the families which Schlechta investigated still do not admit a normal characterization,
in our larger sense. This is therefore a generalization of his negative results.
Second, we will extend the negative results (always in our sense) to new families of distance operators,
in particular to some that respect the Hamming distance.

We are quite confident that the present work can be continued,
like the work of Schlechta, to show that for families of distance-based revision operators,
there is no normal characterization either. But, we will cover more families and with
a more general definition of a normal characterization.
This is the main motivation. In addition,
the impossibility results of the present paper already help to understand more clearly the limits of what is possible in this area.
They have therefore an interest of their own.

First, we will present the distance-based revision
and the characterizations of Lehmann {\it et al.}
Second, we will define formally the normal characterizations.
Third, we will show the impossibility results.
And finally, we will conclude.

\section{Background}\label{DISbackground}

\subsection{Pseudo-distances}\label{DISpseudodist}

In many circumstances, it is reasonable to assume that an agent can evaluate
for any two valuations $v$ and $w$, how far is the situation described by $w$ from the situation
described by $v$, or how difficult or unexpected
the transition from $v$ to $w$ is, etc. 
In \cite{LehmannMagidorSchlechta1}, this is modelled by
pseudo-distances:

\begin{definition}\label{DISdiststruc}
Let $\cal V$ be a set.
\\
${\cal D}$ is a {\it pseudo-distance} on $\cal V$ iff
${\cal D} = \langle {C}, \prec, d \rangle$, where ${C}$ is a non-empty set,
$\prec$ is a strict total order on ${C}$, and $d$ is a function from ${\cal V} \times {\cal V}$ to ${C}$.
\end{definition}
Intuitively, $\cal V$ is a set of valuations.
Each element of ${C}$ represents a ``cost''.
$c \prec c'$ means the cost $c$ is strictly smaller than the cost $c'$.
And, $d(v, w)$ is the cost of the move from $v$ to $w$.
Natural properties that come to mind are those of usual distances.
Before introducing them, we need standard notations:

\begin{notation}
$\cal P$ denotes the power set operator.
\\
For every set $S$, $|S|$ denotes the cardinality of $S$.
\\
$\mathbb{N}$, $\mathbb{N}^{+}$, $\mathbb{R}$, and $\mathbb{R}^+$
denote respectively
the natural, positive natural, real, and positive real numbers.
\\
Let $r \in \mathbb{R}$. Then, $abs(r)$ denotes the absolute value of~$r$.
\\
Let $n, m \in \mathbb{N}$. Then, $[n, m]$ denotes the set of every $k$ in $\mathbb{N}$
(not in $\mathbb{R}$) such that $n \leq k \leq m$.
\end{notation}

\begin{definition}
Suppose ${\cal D} = \langle {C}, \prec, d \rangle$ is a pseudo-distance on a set $\cal V$.
\\
${\cal D}$ is {\it symmetric} iff $\forall \: v, w \in {\cal V}$, $d(v, w) = d(w, v)$.
\\
${\cal D}$ is {\it identity respecting} (IR) iff
\\
$(1)$ $C = \mathbb{R}$;
\\
$(2)$ $\prec$ is the usual strict total order on $\mathbb{R}$;
\\
$(3)$ $\forall \: v, w \in {\cal V}$,  $d(v, w) = 0$ iff $v = w$.
\\
${\cal D}$ is {\it positive} iff $(1)$, $(2)$, and
\\
$(4)$ $\forall \: v, w \in {\cal V}$, $0 \preceq d(v, w)$.
\\
${\cal D}$ is {\it triangle-inequality respecting} (TIR) iff $(1)$, $(2)$, and
\\
$(5)$ $\forall \: v, w, x \in {\cal V}$, $d(v, x) \preceq d(v, w) + d(w,x)$.
\end{definition}
These properties have not been imposed from start because
natural circumstances could then no longer be modelled.
For instance, non-symmetric pseudo-distances are useful
when moving from $v$ to $w$ may be ``cheaper'' than moving from $w$ to $v$.
There are also circumstances where staying the same requires effort and then
non-IR pseudo-distances will be helpful.
We can also imagine scenarios where some costs can be seen as ``benefits'',
we will then turn to non-positive pseudo-distances, etc.

In addition, the costs are not required to be necessarily the real numbers.
Indeed, for instance, we could need $|\mathbb{N}|$ to model an ``infinite cost''
useful when a move is impossible or extremely difficult.
Provided one accepts the infinite cost $|\mathbb{N}|$, we can define naturally ``liberal'' versions of identity respect, positivity, and triangle-inequality respect:

\begin{definition}
Suppose ${\cal D} = \langle {C}, \prec, d \rangle$ is a pseudo-distance on a set $\cal V$.
\\
${\cal D}$ is {\it liberally} IR iff
\\
$(1)$ $C = \mathbb{R} \cup \lbrace |\mathbb{N}| \rbrace$;
\\
$(2)$ $\forall \: c, c' \in C$, $c \prec c'$ iff
$(c, c' \in \mathbb{R}$ and $c < c')$ or $(c \in \mathbb{R}$ and $c' = |\mathbb{N}|)$;
\\
$(3)$ $\forall \: v, w \in {\cal V}$,  $d(v, w) = 0$ iff $v = w$.
\\
${\cal D}$ is {\it liberally positive} iff $(1)$, $(2)$, and
\\
$(4)$ $\forall \: v, w \in {\cal V}$, $0 \preceq d(v, w)$.
\\
${\cal D}$ is {\it liberally} TIR iff $(1)$, $(2)$, and
\\
$(5)$ $\forall \: v, w, x \in {\cal V}$:
if $d(v,x), d(v,w), d(w,x) \in \mathbb{R}$, then $d(v, x) \preceq d(v, w) + d(w,x)$;
\\
if $d(v,x) = |\mathbb{N}|$, then $d(v, w) = |\mathbb{N}|$ or $d(w, x) = |\mathbb{N}|$.
\end{definition}
The Hamming distance between propositional valuations has been considered in \cite{Dalal1} and
investigated further by many authors. Respecting this distance is an
important property. We need before to present the matrices for a propositional language \cite{Urquhart1}:

\begin{definition}
Let ${\cal L} = \langle {\cal A}, {\cal C} \rangle$ be a propositional language
(${\cal A}$ denotes the atoms and ${\cal C}$ the connectives),
let $\cal F$ be the set of all well-formed formulas (wffs) of $\cal L$, and $\forall \: \diamond \in {\cal C}$,
let $n(\diamond)$ be the arity of $\diamond$.
\\
$\cal M$ is a {\it matrix} on $\cal L$ iff
${\cal M} = \langle T, E, f \rangle$, where
$T$ is a set,
$E$ is a non-empty proper subset of $T$,
and $f$ is a function (whose domain is ${\cal C}$)
such that $\forall \: \diamond \in {\cal C}$, $f_\diamond$ (i.e.~$f(\diamond)$) is a function from
$T^{{n(\diamond)}}$ to $T$.
\\
$v$ is a $\cal M$-{\it valuation} iff
$v$ is a function from $\cal F$ to $T$ such that
$\forall \: \diamond \in {\cal C}$,
$\forall \: \alpha_1, \ldots, \alpha_{{n(\diamond)}} \in {\cal F}$,
$v(\diamond(\alpha_1, \ldots, \alpha_{{n(\diamond)}})) = f_\diamond(v(\alpha_1), \ldots, v(\alpha_{{n(\diamond)}}))$.
\end{definition}
Intuitively, $T$ is a set of truth values and $E$ contains all the designated truth values.

\begin{definition}
Let ${\cal L} = \langle {\cal A}, {\cal C} \rangle$ be a propositional language, $\cal M$ a matrix on $\cal L$,
$\cal V$ the set of all $\cal M$-valuations, and
${\cal D} = \langle {C}, \prec, d \rangle$ a pseudo-distance on $\cal V$.
\\
We use the following notation: $\forall \: v, w \in {\cal V}$,
\\
$h(v, w) := \lbrace p \in {\cal A} : v(p) \not= w(p) \rbrace$.
\\
${\cal D}$ is {\it Hamming-inequality respecting} (HIR) iff
$\forall \: v, w, x \in {\cal V}$,
if $|h(v, w)| < |h(v, x)|$, then $d(v, w) \prec d(v, x)$.
\end{definition}
Recall that $h(v, w)$ may be infinite and thus $<$ should be understood as the usual order on the cardinal numbers.

We turn to crucial operators introduced in \cite{LehmannMagidorSchlechta1}.
They are central in the definition of the distance-based revision.
They transform any two sets of valuations $V$ and $W$ into the set
of every element $w$ of $W$ such that a global move from $V$ to $w$ is of minimal cost.
Note that concerning this point, \cite{LehmannMagidorSchlechta1} has its roots in \cite{KatsunoMendelzon1} and especially in \cite{Lewis1}.

\begin{definition}
Suppose ${\cal D} = \langle {C}, \prec, d \rangle$ is a pseudo-distance on a set $\cal V$.
\\
We denote by $|_{\cal D}$ the binary operator on ${\cal P}({\cal V})$ such that
$\forall \: V, W \subseteq {\cal V}$, we have $V |_{\cal D} W =$
$$
\lbrace w \in W : \exists \: v \in V, \forall \: v' \in V, \forall \: w' \in W, d(v, w) \preceq d(v', w') \rbrace.
$$
\end{definition}

\subsection{Distance-based revision operators} \label{DISdistbaseopera}

The ontological commitments endorsed in \cite{LehmannMagidorSchlechta1}
are close to the AGM ones:
a classical propositional language is considered and both epistemic states and new information
are modelled by consistent sets of formulas
(not necessarily deductively closed).

\begin{notation}
We denote by ${\cal L}_c$ some classical propositional language
and by $\vdash_c$, ${\cal V}_c$, $\models_c$, and ${\cal F}_c$ respectively
the classical consequence relation,
valuations, satisfaction relation, and wffs of ${\cal L}_c$.
Let $\Gamma, \Delta \subseteq {\cal F}_c$ and $V \subseteq {\cal V}_c$, then:
\\
$\Gamma \vee \Delta := \lbrace \alpha \vee \beta : \alpha \in \Gamma, \beta \in \Delta \rbrace$;
\\
$\C{\vdash_c}{\Gamma} := \lbrace \alpha \in {\cal F}_c : \Gamma \vdash_c \alpha \rbrace$;
\\
$\M{\Gamma} := \lbrace v \in {\cal V}_c : \forall \: \alpha \in \Gamma, v \models_c \alpha \rbrace$;
\\
$\T{V} := \lbrace \alpha \in {\cal F}_c : V \subseteq \M{\alpha} \rbrace$;
\\
${\bf C} := \lbrace \Gamma \subseteq {\cal F}_c : \C{\vdash_c}{\Gamma} \not= {\cal F}_c \rbrace$;
\\
${\bf D} := \lbrace V \subseteq {\cal V}_c : \exists \: \Gamma \subseteq {\cal F}_c, V = \M{\Gamma} \rbrace$.
\end{notation}
In this classical framework, two new properties for pseudo-distances can be defined.
They convey natural meanings.
Their importance has been put in evidence in \cite{LehmannMagidorSchlechta1}.

\begin{definition}
Let ${\cal D} = \langle {C}, \prec, d \rangle$ be a pseudo-distance on ${\cal V}_c$.
\\
${\cal D}$ is {\it definability preserving} (DP) iff
\\
$\forall \: V, W \in {\bf D}$, $V|_{\cal D}W \in {\bf D}$.
\\
${\cal D}$ is {\it consistency preserving} (CP) iff
\\
$\forall \: V, W \in {\cal P}({\cal V}_c) \setminus \lbrace \emptyset \rbrace$, $V|_{\cal D}W \not=\emptyset$.
\end{definition}
Now, suppose we are given a pseudo-distance $\cal D$ on ${\cal V}_c$.
Then, the revision of a consistent set of formulas $\Gamma$ by a second one $\Delta$ can
be defined naturally as the set of all formulas satisfied in $\M{\Gamma} |_{\cal D} \M{\Delta}$:

\begin{definition}
Let $\star$ be an operator from ${\bf C} \times {\bf C}$ to ${\cal P}({\cal F}_c)$.
\\
We say that $\star$ is a {\it distance-based revision operator} iff there exists
a pseudo-distance ${\cal D}$ on ${\cal V}_c$ such that
$\forall \: \Gamma, \Delta \in {\bf C}$,
$$\Gamma \star \Delta = \T{\M{\Gamma} |_{\cal D} \M{\Delta}}.$$
In addition, if ${\cal D}$ is symmetric, IR, DP etc., then so is $\star$.
\end{definition}
The authors of \cite{LehmannMagidorSchlechta1} rewrote the AGM postulates in their framework as follows.
\\
Suppose $\star$ is an operator from ${\bf C} \times {\bf C}$ to ${\cal P}({\cal F}_c)$
Then, define the following properties: $\forall \: \Gamma, \Gamma', \Delta, \Delta' \in {\bf C}$,
\begin{description}
\item[$(\star0)$] if $\C{\vdash_c}{\Gamma} = \C{\vdash_c}{\Gamma'}$ and $\C{\vdash_c}{\Delta} = \C{\vdash_c}{\Delta'}$,
\\
then $\Gamma \star \Delta = \Gamma' \star \Delta'$;
\item[$(\star1)$] $\Gamma \star \Delta \in {\bf C}$ and $\Gamma \star \Delta = \C{\vdash_c}{\Gamma \star \Delta}$;
\item[$(\star2)$] $\Delta \subseteq \Gamma \star \Delta$;
\item[$(\star3)$] if $\Gamma \cup \Delta \in {\bf C}$, then $\Gamma \star \Delta = \C{\vdash_c}{\Gamma \cup \Delta}$;
\item[$(\star4)$] if $(\Gamma \star \Delta) \cup \Delta' \in {\bf C}$,
\\
then
$\Gamma \star (\Delta \cup \Delta') = \C{\vdash_c}{(\Gamma \star \Delta) \cup \Delta'}$.
\end{description}
Then, it can be checked that every positive, IR, CP and DP distance-based revision operator~$\star$
satisfies $(\star0)$-$(\star4)$, i.e. the AGM postulates.
More importantly, $\star$ satisfies also certain properties that deal with iterated revisions.
This is not surprising as the revisions of the different $\Gamma$'s are all defined by a
unique pseudo-distance, which ensures a strong coherence between them.
For example, $\star$ satisfies two following properties:
$\forall \: \Gamma, \Delta, \lbrace \alpha \rbrace, \lbrace \beta \rbrace \in {\bf C}$,
\begin{itemize}
\item if $\gamma \in (\Gamma \star \lbrace \alpha \rbrace) \star \Delta$ and $\gamma \in (\Gamma \star \lbrace \beta \rbrace) \star \Delta$,
\\
then $\gamma \in (\Gamma \star \lbrace \alpha \vee \beta \rbrace) \star \Delta$;
\item if $\gamma \in (\Gamma \star \lbrace \alpha \vee \beta \rbrace) \star \Delta$,
\\
then
$\gamma \in (\Gamma \star \lbrace \alpha \rbrace) \star \Delta$
or $\gamma \in (\Gamma \star \lbrace \beta \rbrace) \star \Delta$.
\end{itemize}
These properties are not entailed by the AGM postulates,
a counter-example can be found in \cite{LehmannMagidorSchlechta1}.
But, they seem intuitively justified.
Indeed, take three sequences of revisions that differ only at some step in which
the new information is $\alpha$ in the first sequence, $\beta$ in the second,
and $\alpha \vee \beta$ in the third.
Now, suppose $\gamma$ is concluded after both the first and the second sequences.
Then, it should intuitively be the case that $\gamma$ is concluded after the third sequence too.
Similar arguments can be given for the second property.
Now, to characterize the full distance-based revision more is needed.
This is discussed in the next section.

\subsection{Characterizations}\label{DIScharacLMS}

The authors of \cite{LehmannMagidorSchlechta1} provided characterizations
for families of distance-based revision operators.
They proceed in two steps. First, they defined the distance operators,
in a very general framework:

\begin{definition}
Let $\cal V$ be a set, ${\bf V}, {\bf W}, {\bf X} \subseteq {\cal P}({\cal V})$, and $|$ an operator
from ${\bf V} \times {\bf W}$ to ${\bf X}$.
\\
$|$ is a {\it distance operator} iff there exists a pseudo-distance ${\cal D}$ on $\cal V$ such that
$\forall \: V \in {\bf V}$, $\forall \: W \in {\bf W}$, $V | W = V |_{\cal D} W$.
\\
In addition, if ${\cal D}$ is symmetric, HIR, DP, etc., then so is $|$.
\end{definition}
Then, they characterized families of such distance operators
(with the least possible assumptions about $\bf V$, $\bf W$, and $\bf X$). This is the essence of their work.
Here is an example: 

\begin{proposition}\label{DISalgebraic}{\bf \cite{LehmannMagidorSchlechta1}}
\\
Suppose ${\cal V}$ is a non-empty set, ${\bf V} \subseteq {\cal P}({\cal V})$
(such that $\emptyset \not\in {\bf V}$ and $\forall \: V, W \in {\bf V}$,
we have $V \cup W \in {\bf V}$ and if $V \cap W \not=\emptyset$, then $V \cap W \in {\bf V}$ too),
and $|$ an operator from ${\bf V} \times {\bf V}$ to ${\bf V}$.
\\
Then, $|$ is a symmetric distance operator iff $\forall \: k \in \mathbb{N}^+$ and
$\forall \: V_0, V_1, \ldots, V_k \in {\bf V}$, we have $V_0|V_1 \subseteq V_1$ and
\begin{description}
\item[$(| loop)$] if $\left \{ \begin{array}{l} (V_1 | (V_0 \cup V_2)) \cap V_0 \not= \emptyset,\\
(V_2 | (V_1 \cup V_3)) \cap V_1 \not= \emptyset,\\
\ldots,\\
(V_k | (V_{k-1} \cup V_0)) \cap V_{k-1} \not= \emptyset, \end{array} \right .$
\\
then
$(V_0 | (V_k \cup V_1)) \cap V_1 \not= \emptyset$.
\end{description}
\end{proposition}
In a second step only, they applied these results to characterize families of distance-based revision operators.
For instance, they applied Proposition~\ref{DISalgebraic} to get Proposition~\ref{DIScaracrevclass} below.
We should say immediately that they chose a classical framework to
define the distance-based revision.
But, if we choose now another framework, there are quite good chances
that Proposition~\ref{DISalgebraic} can be still applied, thanks to its algebraic nature.

\begin{proposition}\label{DIScaracrevclass} {\bf \cite{LehmannMagidorSchlechta1}}
\\
Let $\star$ be an operator from ${\bf C} \times {\bf C}$ to~${\cal P}({\cal F}_c)$.
\\
Then, $\star$ is a symmetric CP DP distance-based revision operator iff
$\star$ satisfies $(\star0)$, $(\star1)$, $(\star2)$, and
\\
$\forall \: k \in \mathbb{N}^+$, $\forall \:
\Gamma_0, \Gamma_1, \ldots, \Gamma_k \in {\bf C}$,
\begin{description}
\item[$(\star loop)$] if $\left \{ \begin{array}{l}
\Gamma_0 \cup (\Gamma_1 \star (\Gamma_0 \vee \Gamma_2)) \in {\bf C},\\
\Gamma_1 \cup (\Gamma_2 \star (\Gamma_1 \vee \Gamma_3)) \in {\bf C},\\
\ldots,\\
\Gamma_{k-1} \cup (\Gamma_k \star (\Gamma_{k-1} \vee \Gamma_0)) \in {\bf C},
\end{array} \right .$
\\
then
$\Gamma_1 \cup (\Gamma_0 \star (\Gamma_k \vee \Gamma_1)) \in {\bf C}$.
\end{description}
\end{proposition}

\section{Normal characterizations}\label{DISnormalchar}

Let $\cal V$ be a set, ${\cal O}$ a set of binary operators on ${\cal P}({\cal V})$,
and $|$ a binary operator on ${\cal P}({\cal V})$.
Approximatively, in \cite{Schlechta5},
a characterization of ${\cal O}$ is called normal iff it contains only conditions which are
universally quantified, apply $|$ only a finite number of times, and use only elementary operations
(like~e.g.~$\cup$, $\cap$, $\setminus$), see Section~1.6.2.1 of \cite{Schlechta5} for details.
Here is an example of such a condition:
\begin{description}
\item[$(C1)$] $\forall \: V, W \in {\bf U} \subseteq {\cal P}({\cal V})$, $V | ( (V \cup W) | W ) = \emptyset$.
\end{description}
%

Now, we introduce a new, more general, definition with an aim of
providing more general impossibility results.
Approximatively, in the present paper, a characterization of $\cal O$ will be called normal iff
it contains only conditions which are universally quantified and apply $|$ only a finite number of times.
Then, the conditions can involve complex structures or functions, etc.,
we are not limited to elementary operations. More formally:

\begin{definition} \label{DISnormalCO}
Suppose $\cal V$ is a set and ${\cal O}$ a set of binary operators on ${\cal P}({\cal V})$.
\\
$\cal C$ is a {\it normal characterization} of ${\cal O}$ iff
${\cal C} = \langle n, \Phi \rangle$ where $n \in \mathbb{N}^+$ and $\Phi$ is a relation on
${\cal P}({\cal V})^{3n}$ such that for every binary operator $|$ on ${\cal P}({\cal V})$, we have
$| \in {\cal O}$ iff
\\
$\forall \: V_1, \ldots, V_n, W_1, \ldots, W_n \subseteq {\cal V}$,
\\
$(V_1, \ldots, V_n, W_1, \ldots, W_n,
V_1 | W_1, \ldots, V_n | W_n) \in \Phi.$
\end{definition}
Note that $\Phi$ is a relation in the purely set-theoretic sense.
Now, suppose there is no normal characterization of ${\cal O}$.
Here are examples (i.e. $(C1)$, $(C2)$, and $(C3)$ below) that will give the reader a good idea
which conditions cannot characterize ${\cal O}$.
This will therefore make clearer the range
of our impossibility results (Propositions~\ref{DISpascarac} and \ref{DISpascaracham} below).

To begin, $(C1)$ cannot characterize $\cal O$.
Indeed, suppose it does, i.e.
$| \in {\cal O}$ iff $\forall \: V, W \in {\bf U}$, $V | ( (V \cup W) | W ) = \emptyset$.
\\
Then, take $n = 4$ and $\Phi$ such that
\\
$(V_1, \dots, V_4, W_1, \ldots, W_4, X_1, \ldots, X_4) \in \Phi$ iff
\\
$\left \{ \begin{array}{l} V_1, V_2 \in {\bf U},\\
V_3 = V_1 \cup V_2,\\
W_3 = V_2,\\
V_4 = V_1,\\
W_4 = X_3\end{array} \right .$
entail $X_4 = \emptyset$.
\\
Then, $\langle 4, \Phi \rangle$ is a normal characterization of ${\cal O}$.
We give the easy proof of this, so that the reader can check 
that a convenient relation $\Phi$ can be found immediately for
all simple conditions like $(C1)$.

\begin{proof}
Direction: ``$\rightarrow$''.
\\
Suppose $| \in {\cal O}$.
\\
Then, $\forall \: V, W \in {\bf U}$, $V | ( (V \cup W) | W ) = \emptyset$.
\\
Let $V_1, \ldots, V_4, W_1, \ldots, W_4 \subseteq {\cal V}$. We show:
\\
$(V_1, \ldots, V_4, W_1, \ldots, W_4,
V_1 | W_1, \ldots, V_4 | W_4) \in \Phi$.
\\
Suppose $V_1, V_2 \in {\bf U}$, $V_3 = V_1 \cup V_2$, $W_3 = V_2$, $V_4 = V_1$, and $W_4 = V_3 | W_3$.
\\
Then, as $V_1, V_2 \in {\bf U}$, we get $V_1 | ( (V_1 \cup V_2) | V_2 ) = \emptyset$.
\\
But, $V_1 | ( (V_1 \cup V_2) | V_2 ) = V_1 | ( V_3 | W_3 ) = V_4 | W_4$.

Direction: ``$\leftarrow$''.
\\
Suppose $\forall \: V_1, \ldots, V_4, W_1, \ldots, W_4 \subseteq {\cal V}$,
\\
$(V_1, \ldots, V_4, W_1, \ldots, W_4, V_1 | W_1, \ldots, V_4 | W_4) \in \Phi$.
\\
We show $| \in {\cal O}$.
Let $V, W \in {\bf U}$.
\\
Take $V_1 = V$,
$V_2 = W$,
$V_3 = V_1 \cup V_2$,
$W_3 = V_2$,
$V_4 = V_1$,
$W_4 = V_3 | W_3$.
Take any values for $W_1$ and $W_2$.
\\
Then, $V_1 \in {\bf U}$, $V_2 \in {\bf U}$, $V_3 = V_1 \cup V_2$, $W_3 = V_2$, $V_4 = V_1$, and $W_4 = V_3|W_3$.
\\
But, $(V_1, \ldots, V_4, W_1, \ldots, W_4, V_1 | W_1, \ldots, V_4 | W_4) \in \Phi$
\\
Therefore, by definition of $\Phi$, $V_4 | W_4 = \emptyset$.
\\
But, $V_4 | W_4 =
V_1 | ((V_1 \cup V_2) | V_2) = V | ((V \cup W) | W)$.\qed
\end{proof}
At this point, we excluded all those conditions which are excluded
by (the nonexistence of a normal characterization of ${\cal O}$ in the sense of) Schlecha,
i.e. all conditions like $(C1)$.
But actually, more complex conditions are also excluded.
For instance, let $f$ be any function from ${\cal P}({\cal V})$ to ${\cal P}({\cal V})$.
Then, the following condition:
\begin{description}
\item[$(C2)$] $\forall \: V, W \in {\bf U}$, $f(V) | ( (V \cup W) | W ) = \emptyset$.
\end{description}
cannot characterize ${\cal O}$. Indeed, suppose it characterizes ${\cal O}$.
Then, take $n = 4$ and $\Phi$ such that
\\
$(V_1, \dots, V_4, W_1, \ldots, W_4, X_1, \ldots, X_4) \in \Phi$ iff
\\
$\left \{ \begin{array}{l}
V_1, V_2 \in {\bf U},\\
V_3 = V_1 \cup V_2,\\
W_3 = V_2,\\
V_4 = f(V_1),\\
W_4 = X_3 \end{array} \right .$
entail $X_4 = \emptyset$.
\\
Then, $\langle 4, \Phi \rangle$ is a normal characterization of ${\cal O}$.
We leave the easy proof of this to the reader.
On the other hand, $(C2)$ is not excluded by Schlechta,
if $f$ cannot be constructed from elementary operations.
But, even if there exists such a construction,
showing that it is indeed the case might well be a difficult problem. 

We can even go further combining universal (not existential) quantifiers and functions like~$f$.
For instance, suppose ${\cal G}$ is a set of functions from ${\cal P}({\cal V})$ to ${\cal P}({\cal V})$ and
consider the following condition:
\begin{description}
\item[$(C3)$] $\forall \: f \in {\cal G}$, $\forall \: V, W \in {\bf U}$, $f(V) | ( (V \cup W) | W ) = \emptyset$.
\end{description}
Then, $(C3)$ cannot characterize ${\cal O}$. Indeed, suppose $(C3)$ characterizes ${\cal O}$.
Then, take $n = 4$ and $\Phi$ such that
\\
$(V_1, \dots, V_4, W_1, \ldots, W_4, X_1, \ldots, X_4) \in \Phi$ iff
\\
$\forall \: f \in {\cal G}$, if $\left \{ \begin{array}{l}
V_1, V_2 \in {\bf U},\\
V_3 = V_1 \cup V_2,\\
W_3 = V_2,\\
V_4 = f(V_1),\\
W_4 = X_3, \end{array} \right .$ then $X_4 = \emptyset$.
\\
Then, $\langle 4, \Phi \rangle$ is a normal characterization of ${\cal O}$.
The easy proof is left to the reader. On the other hand, $(C3)$ is not excluded by Schlechta.

Finally, a good example of a condition which is not excluded (neither by us nor by Schlechta)
is of course the arbitrary big loop condition $(| loop)$.

\section{Impossibility results}\label{DISnoFiniteNormChar}

We provide our first impossibility result.
It generalizes Proposition 4.2.11 of \cite{Schlechta5}.
Our proof will be based on a slight adaptation of
a particular pseudo-distance invented by Schlechta (called ``Hamster Wheel'').

\begin{proposition}\label{DISpascarac}
Let $\cal V$ be an infinite set, ${\cal N}$ the set of all symmetric IR positive TIR distance operators
from ${\cal P}({\cal V})^2$ to ${\cal P}({\cal V})$, and
${\cal O}$ a set of distance operators from ${\cal P}({\cal V})^2$ to ${\cal P}({\cal V})$
such that ${\cal N} \subseteq {\cal O}$.
\\
Then, there does not exist a normal characterization of ${\cal O}$.
\end{proposition}

\begin{proof}
Suppose the contrary, i.e. suppose there is $n \in \mathbb{N}^+$ and
a relation $\Phi$ on ${\cal P}({\cal V})^{3n}$ such that
\begin{description}
\item[$(0)$] for every binary operator $|$ on ${\cal P}({\cal V})$,
we have $| \in {\cal O}$ iff
\\
$\forall \: V_1, \ldots, V_n$, $W_1, \ldots, W_n \subseteq {\cal V}$,
\\
$(V_1, \ldots, V_n$, $W_1, \ldots, W_n$, $V_1 | W_1, \ldots, V_n | W_n) \in \Phi$.
\end{description}
As $\cal V$ is infinite, there are distinct $v_1, \ldots,  v_m$, $w_1, \ldots,  w_m$ in ${\cal V}$, with $m = n + 3$.
\\
Let $X = \lbrace v_1, \ldots,  v_m, w_1, \ldots,  w_m \rbrace$.
\\
Let ${\cal D}$ be the pseudo-distance on $\cal V$ such that
${\cal D} = \langle \mathbb{R}, <, d \rangle$, where $<$ is the usual order on $\mathbb{R}$ and
$d$ is the function defined as follows. Let $v, w \in {\cal V}$. Consider the cases that follow:
\\
Case~1: $v = w$.
\\
Case~2: $v \not= w$.
\\
Case~2.1: $\lbrace v, w \rbrace \not\subseteq X$.
\\
Case~2.2: $\lbrace v, w \rbrace \subseteq X$.
\\
Case~2.2.1: $\lbrace v, w \rbrace \subseteq \lbrace v_1, \ldots, v_m \rbrace$.
\\
Case~2.2.2: $\lbrace v, w \rbrace \subseteq \lbrace w_1, \ldots, w_m \rbrace$.
\\
Case~2.2.3: $\exists \: i, j \in [1, m],\; \lbrace v, w \rbrace = \lbrace v_i, w_j \rbrace$.
\\
Case~2.2.3.1: $i = j$.
\\
Case~2.2.3.2: $abs(i - j) \in \lbrace 1, m-1 \rbrace$.
\\
Case~2.2.3.3: $1 < abs(i - j) < m-1$.
\\
Then,
$$
d(v, w) = \left \{ \begin{array}{l l}
0 & \textrm{if Case~1 holds;}\\
1 & \textrm{if Case~2.1 holds;}\\
1.1 & \textrm{if Case~2.2.1 holds;} \\
1.1 & \textrm{if Case~2.2.2 holds;} \\
1.4 & \textrm{if Case~2.2.3.1 holds;}\\
2 & \textrm{if Case~2.2.3.2 holds;}\\
1.2 & \textrm{if Case~2.2.3.3 holds.}\\
\end{array} \right .
$$
Note that ${\cal D}$ is essentially, but not exactly, the Hamster Wheel of \cite{Schlechta5}.
The main difference is Case~2.1, which
was not treated by Schlechta. The reader can find a picture of $\cal D$ in Figure~1.
\\
\begin{center}
\epsfig{file=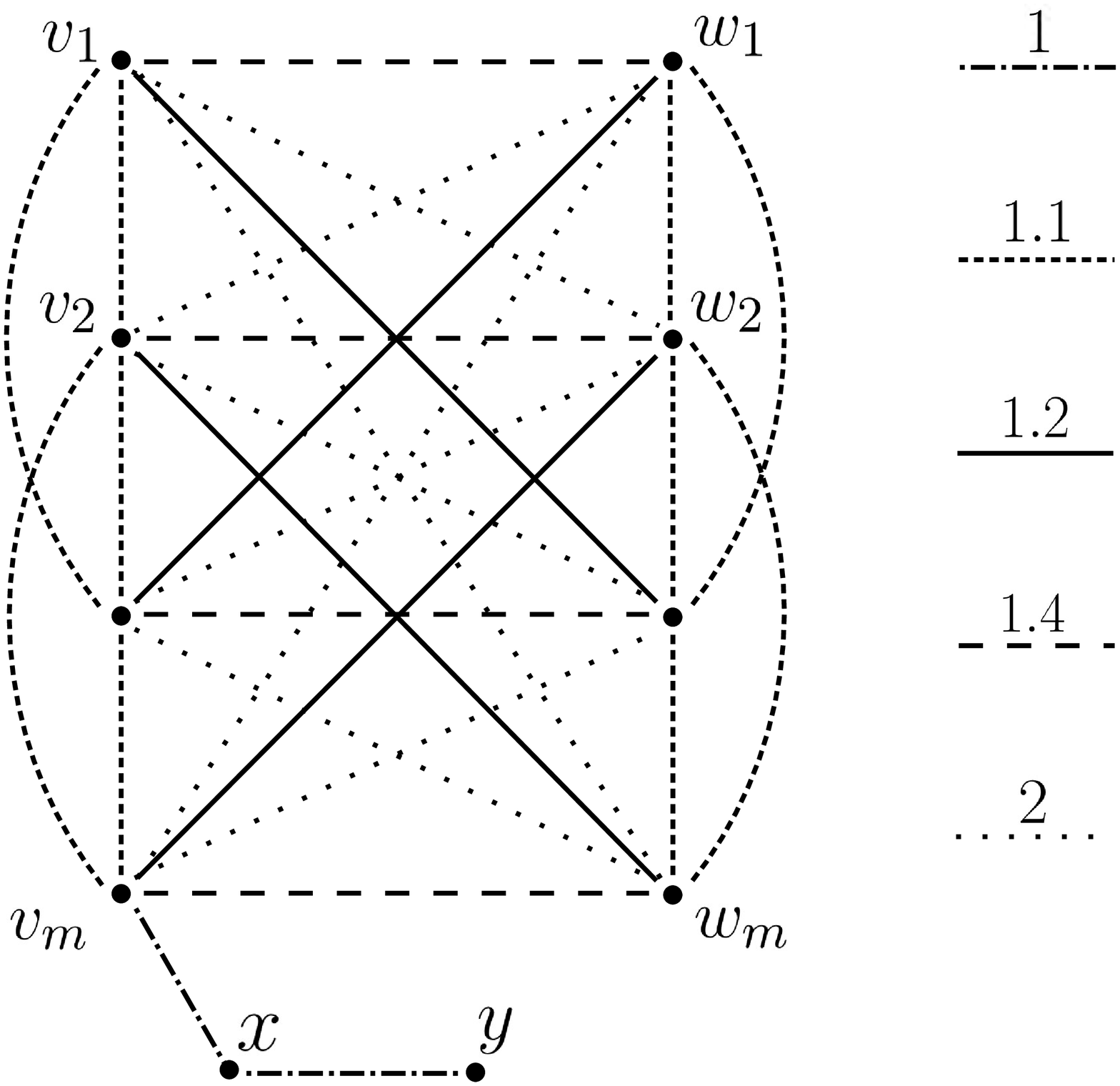, height=8cm}
\hfill \\
Figure 1: A slight adaptation of Hamster Wheel.
\end{center}
\hfill \\
Let $|$ be the binary operator on ${\cal P}({\cal V})$ such that $\forall \: V, W \subseteq {\cal V}$,
$$
V | W = \left \{ \begin{array}{l l}
\lbrace w_m \rbrace & \textrm{if}\; V = \lbrace v_m, v_1 \rbrace, W = \lbrace w_m, w_1 \rbrace;\\
\lbrace v_m \rbrace & \textrm{if}\; V = \lbrace w_m, w_1 \rbrace, W = \lbrace v_m, v_1 \rbrace;\\
V |_{\cal D} W & \textrm{otherwise.}
\end{array} \right .
$$
The difference between $|$ and $|_{\cal D}$ is strong enough so that:
\begin{description}
\item[$(1)$] $|$ is not a distance operator.
\end{description}
The proof will be given later.
Thus, $| \not\in {\cal O}$. Thus, by $(0)$:
\begin{description}
\item[$(2)$] $\exists V_1, \ldots, V_n, W_1, \ldots, W_n \subseteq {\cal V}$,
\\
$(V_1, \ldots, V_n, W_1, \ldots, W_n, V_1 | W_1, \ldots, V_n | W_n) \not\in \Phi$.
\end{description}
In addition, we took $m$ sufficiently big so that:
\begin{description}
\item[$(3)$] $\exists \: r \in [1, m-1]$ such that
\\
$\forall \: i \in [1, n]$, $\lbrace V_i, W_i \rbrace \not= \lbrace \lbrace v_r, v_{r+1} \rbrace, \lbrace w_r, w_{r+1} \rbrace \rbrace$.
\end{description}
We will give the proof later.
\\
Let $|'$ be the binary operator on ${\cal P}({\cal V})$ such that $\forall \: V, W \subseteq {\cal V}$,
$$
V |' W = \left \{ \begin{array}{l l}
\lbrace w_{r+1} \rbrace & \!\!\!\textrm{if}\: V = \lbrace v_r, v_{r+1} \rbrace, W = \lbrace w_r, w_{r+1} \rbrace;\\
\lbrace v_{r+1} \rbrace & \!\!\!\textrm{if}\: V = \lbrace w_r, w_{r+1} \rbrace, W = \lbrace v_r, v_{r+1} \rbrace;\\
V | W & \!\!\!\textrm{otherwise.}\\
\end{array} \right .
$$
The difference between $|'$ and $|$ is ``invisible''
for $\Phi$.
\\
More formally,
$\forall \: i \in [1, n]$, $V_i |' W_i = V_i | W_i$.
\\
The proof of this is obvious by $(3)$.
\\
Therefore, by $(2)$, we get:
\\
$(V_1, \ldots, V_n$, $W_1, \ldots, W_n$, $V_1 |' W_1, \ldots, V_n |' W_n) \not\in \Phi$.
\\
Thus, by $(0)$, we obtain:
\begin{description}
\item[$(4)$] $|' \not\in {\cal O}$.
\end{description}
But, at the same time, there is a convenient pseudo-distance that represents $|'$.
Indeed, let ${\cal D}'$ be the
pseudo-distance on $\cal V$ such that ${\cal D}' = \langle \mathbb{R}, <, d' \rangle$, where
$d'$ is the function such that $\forall \: v, w \in {\cal V}$,
$$
d'(v, w) = \left \{ \begin{array}{l l}
\!\!\!1.3 & \!\!\!\textrm{if}\: \exists i \in [r+1, m], \lbrace v, w \rbrace = \lbrace v_i, w_i \rbrace; \\
\!\!\!d(v, w) & \!\!\!\textrm{otherwise.}
\end{array} \right .
$$
Then, we will show:
\begin{description}
\item[$(5)$] $|' = |_{{\cal D}'}$.
\end{description}
But, ${\cal D}'$ is obviously symmetric, IR, and positive.
\\
In addition, ${\cal D}'$ is TIR, because ${\cal D}'$ is IR and
\\
$\forall \: v, w \in {\cal V}$, $d'(v, w) = 0$ or $1 \leq d'(v,w) \leq 2$.
\\
Thus, $|'$ is a symmetric IR positive TIR distance operator.
\\
Consequently, $|' \in {\cal N}$ and thus
\begin{description}
\item[$(6)$] $|' \in {\cal O}$.
\end{description}
So, we get a final contradiction by $(4)$ and $(6)$.
\\
\\
{\it Proof of $(1)$}. Suppose the contrary, i.e.
suppose there is a pseudo-distance
${{\cal S}} = \langle {C}, \prec, g \rangle$ on $\cal V$ such that $| = |_{{\cal S}}$.
\\
Then, we will show:
\\
$(1.1)$\quad $\forall \: i \in [1, m-1]$, $g(v_i, w_i) = g(v_{i+1}, w_{i+1})$.
\\
On the other hand, we will show:
\\
$(1.2)$\quad $g(v_m, w_m) \prec g(v_1, w_1)$.
\\
But, by $(1.1)$ and $(1.2)$, we get an obvious contradiction.
\\
\\
{\it Proof of $(1.1)$}. 
Suppose $i \in [1, m-1]$.
Then:
\\
$\lbrace v_i, v_{i+1} \rbrace |_{{\cal S}}\lbrace w_i, w_{i+1} \rbrace
= \lbrace v_i, v_{i+1} \rbrace |_{\cal D} \lbrace w_i, w_{i+1} \rbrace
= \lbrace w_i, w_{i+1} \rbrace$.
\\
Case~1: $g(v_i, w_i) \prec g(v_{i+1}, w_{i+1})$.
\\
We have $\lbrace v_i \rbrace |_{{\cal S}}\lbrace w_i, w_{i+1} \rbrace =
\lbrace v_i \rbrace |_{\cal D} \lbrace w_i, w_{i+1} \rbrace = \lbrace w_i \rbrace$.
\\
Thus, $w_{i+1} \not\in \lbrace v_i \rbrace |_{{\cal S}}\lbrace w_i, w_{i+1} \rbrace$.
\\
Therefore, $g(v_i, w_i) \prec g(v_i, w_{i+1})$.
\\
Thus, $w_{i+1} \not\in \lbrace v_i, v_{i+1} \rbrace |_{{\cal S}}\lbrace w_i, w_{i+1} \rbrace$,
which is impossible.
\\
Case~2: $g(v_{i+1}, w_{i+1}) \prec g(v_i, w_i)$.
\\
We have $\lbrace v_{i+1} \rbrace |_{{\cal S}}\lbrace w_i, w_{i+1} \rbrace =
\lbrace v_{i+1} \rbrace |_{\cal D} \lbrace w_i, w_{i+1} \rbrace = \lbrace w_{i+1} \rbrace$.
\\
Therefore, $w_i \not\in \lbrace v_{i+1} \rbrace |_{{\cal S}}\lbrace w_i, w_{i+1} \rbrace$.
\\
Consequently, $g(v_{i+1}, w_{i+1}) \prec g(v_{i+1}, w_i)$.
\\
Thus, $w_i \not\in \lbrace v_i, v_{i+1} \rbrace |_{{\cal S}}\lbrace w_i, w_{i+1} \rbrace$,
which is impossible.
\\
Case~3: $g(v_i, w_i) \not\prec g(v_{i+1}, w_{i+1})$ and $g(v_{i+1}, w_{i+1}) \not\prec g(v_i, w_i)$.
\\
Then, as $\prec$ is total, $g(v_i, w_i) = g(v_{i+1}, w_{i+1})$.
\\
\\
{\it Proof of $(1.2)$}.
We have $\lbrace v_m, v_1 \rbrace |_{{\cal S}} \lbrace w_m, w_1 \rbrace =$ $\lbrace v_m, v_1 \rbrace | \lbrace w_m, w_1 \rbrace$ $=$ $\lbrace w_m \rbrace$.
\\
Therefore, $w_1 \not\in \lbrace v_m, v_1 \rbrace |_{{\cal S}} \lbrace w_m, w_1 \rbrace$.
Thus:
\\
$\exists \: v \in \lbrace v_m, v_1 \rbrace$, $\exists \: w \in \lbrace w_m, w_1 \rbrace$,
$g(v, w) \prec g(v_1, w_1)$.
\\
Case~1: $g(v_m, w_m) \prec g(v_1, w_1)$. We are done.
\\
Case~2: $g(v_m, w_1) \prec g(v_1, w_1)$.
\\
We have $\lbrace v_m \rbrace |_{{\cal S}}\lbrace w_m, w_1 \rbrace =
\lbrace v_m \rbrace |_{\cal D} \lbrace w_m, w_1 \rbrace = \lbrace w_m \rbrace$.
\\
Therefore, $w_1 \not\in \lbrace v_m \rbrace |_{{\cal S}}\lbrace w_m, w_1 \rbrace$.
\\
Thus, $g(v_m, w_m) \prec g(v_m, w_1)$.
\\
Thus, by transitivity of $\prec$, $g(v_m, w_m) \prec g(v_1, w_1)$.
\\
Case~3: $g(v_1, w_m) \prec g(v_1, w_1)$.
\\
Then, $\lbrace v_1 \rbrace |_{{\cal S}}\lbrace w_m, w_1 \rbrace = \lbrace w_m \rbrace$.
\\
However, $\lbrace v_1 \rbrace |_{{\cal S}}\lbrace w_m, w_1 \rbrace =
\lbrace v_1 \rbrace |_{\cal D} \lbrace w_m, w_1 \rbrace = \lbrace w_1 \rbrace$,
which is impossible.
\\
Case~4: $g(v_1, w_1) \prec g(v_1, w_1)$.
\\
Impossible by irreflexivity of $\prec$.
\\
\\
{\it Proof of $(3)$}. For all $s \in [1, m-1]$, define:
\\
$I_s := \lbrace i \in [1, n] : \lbrace V_i, W_i \rbrace = \lbrace \lbrace v_s, v_{s+1} \rbrace,
\lbrace w_s, w_{s+1} \rbrace \rbrace\rbrace$.
\\
Suppose the opposite of what we want to show, i.e. suppose $\forall \: s \in [1, m-1]$,
$I_s \not = \emptyset$.
\\
As $v_1, \ldots,  v_m, w_1, \ldots,  w_m$ are distinct, $\forall \: s, t \in [1, m-1]$, if $s \not= t$, then
$I_s \cap I_t = \emptyset$.
\\
Therefore, $m-1 \leq | I_1 \cup \ldots \cup I_{m-1}|$.
\\
On the other hand, $\forall \: s \in [1, m-1]$, $I_s \subseteq [1, n]$.
\\
Thus, $| I_1 \cup \ldots \cup I_{m-1}| \leq n$.
\\
Thus, $m-1 \leq n$, which is impossible as $m = n+3$.
\\
\\
{\it Proof of $(5)$}. Let $V, W \subseteq {\cal V}$.
\\
Case~1: $V = \lbrace v_r, v_{r+1} \rbrace$ and $W = \lbrace w_r, w_{r+1} \rbrace$.
\\
Then, $V |' W = \lbrace w_{r+1} \rbrace = V |_{{\cal D}'} W$.
\\
Case~2: $V = \lbrace w_r, w_{r+1} \rbrace$ and $W = \lbrace v_r, v_{r+1} \rbrace$.
\\
Then, $V |' W = \lbrace v_{r+1} \rbrace = V |_{{\cal D}'} W$.
\\
Case~3: $V = \lbrace v_m, v_1 \rbrace$ and $W = \lbrace w_m, w_1 \rbrace$.
\\
Then, $V |' W = V | W = \lbrace w_m \rbrace = V |_{{\cal D}'} W$.
\\
Case~4: $V = \lbrace w_m, w_1 \rbrace$ and $W = \lbrace v_m, v_1 \rbrace$.
\\
Then, $V |' W = V | W = \lbrace v_m \rbrace = V |_{{\cal D}'} W$.
\\
Case~5: $\lbrace V, W \rbrace \not\in$
\\
$\lbrace \lbrace \lbrace v_r, v_{r+1} \rbrace, \lbrace w_r, w_{r+1} \rbrace \rbrace,
\lbrace \lbrace v_m, v_1 \rbrace, \lbrace w_m, w_1 \rbrace \rbrace \rbrace$.
\\
Then, $V |' W = V | W = V |_{\cal D} W$.
\\
Case~5.1: $V = \emptyset$ or $W = \emptyset$.
\\
Then, $V |_{\cal D} W = \emptyset = V |_{{\cal D}'} W$.
\\
Case~5.2: $V \cap W \not= \emptyset$.
\\
Then, $V |_{\cal D} W = V \cap W = V |_{{\cal D}'} W$.
\\
Case~5.3: $V \not= \emptyset$, $W \not= \emptyset$, and $V \cap W = \emptyset$.
\\
Case~5.3.1: $V \not\subseteq X$.
\\
Then, $V |_{\cal D} W = W = V |_{{\cal D}'} W$.
\\
Case~5.3.2: $V \subseteq X$.
\\
Case~5.3.2.1: $W \not\subseteq X$.
\\
Then, $V |_{\cal D} W = W \setminus X = V |_{{\cal D}'} W$.
\\
Case~5.3.2.2: $W \subseteq X$.
\\
Case~5.3.2.2.1: $V \not\subseteq \lbrace v_1, \ldots, v_m \rbrace$ and $V \not\subseteq \lbrace w_1, \ldots, w_m \rbrace$.
\\
Then, $V |_{\cal D} W = W = V |_{{\cal D}'} W$.
\\
Case~5.3.2.2.2: $V \subseteq \lbrace v_1, \ldots v_m \rbrace$ and $W \not\subseteq \lbrace w_1, \ldots w_m \rbrace$.
\\
Then, $V |_{\cal D} W = W \cap \lbrace v_1, \ldots, v_m \rbrace = V |_{{\cal D}'} W$.
\\
Case~5.3.2.2.3: $V \subseteq \lbrace v_1, \ldots v_m \rbrace$ and $W \subseteq \lbrace w_1, \ldots w_m \rbrace$.
\\
Case~5.3.2.2.3.1: $\exists \: v_i \in V$, $\exists \: w_j \in W$,
\\
$1 < abs(i-j) < m-1$.
\\
Then, $V |_{\cal D} W =$
\\
$\lbrace w_j \in W : \exists \: v_i \in V$, $1 < abs(i-j) < m-1 \rbrace = V |_{{\cal D}'} W$.
\\
Case~5.3.2.2.3.2: $\forall \: v_i \in V$, $\forall \: w_j \in W$,
\\
$abs(i-j) \in \lbrace 0, 1, m-1 \rbrace$.
\\
Case~5.3.2.2.3.2.1: $|V \cup W| \geq 5$.
\\ 
As $m \geq 4$, $\exists \: v_i \in V$, $\exists \: w_j \in W$, $1 < abs(i-j) < m-1$,
which is impossible.
\\
Case~5.3.2.2.3.2.2: $|V \cup W| \in \lbrace 2, 3, 4 \rbrace$.
\\
Case~5.3.2.2.3.2.2.1: $\lbrace k \in [1, m] : v_k \in V, w_k \in W \rbrace = \emptyset$.
\\
Then, $V |_{\cal D} W = W = V |_{{\cal D}'} W$.
\\
Case~5.3.2.2.3.2.2.2: $\exists \: i \in [1, m]$ such that
\\
$\lbrace k \in [1, m] : v_k \in V, w_k \in W \rbrace = \lbrace i \rbrace$.
\\
Then, $V |_{\cal D} W = \lbrace w_i \rbrace = V |_{{\cal D}'} W$.
\\
Case~5.3.2.2.3.2.2.3: $\exists \: i, j \in [1, m]$ such that $i < j$ and
\\
$\lbrace k \in [1, m] : v_k \in V$ and $w_k \in W \rbrace = \lbrace i, j \rbrace$.
\\
Then, $V = \lbrace v_i, v_j \rbrace$ and $W = \lbrace w_i, w_j \rbrace$.
\\
Case~5.3.2.2.3.2.2.3.1: $r < i$ or $j \leq r$.
\\
Then, $V |_{\cal D} W = \lbrace w_i, w_j \rbrace = V |_{{\cal D}'} W$.
\\
Case~5.3.2.2.3.2.2.3.2: $i \leq r < j$.
\\
We have $abs(i - j) \in \lbrace 1, m-1 \rbrace$.
Thus, $\langle V,\: W \rangle \in$
\\
$\lbrace \langle \lbrace v_r,\: v_{r+1} \rbrace,\: \lbrace w_r,\: w_{r+1} \rbrace \rangle$,
$\langle \lbrace v_1,\: v_m \rbrace,\: \lbrace w_1,\: w_m \rbrace \rangle \rbrace$,
\\
which is impossible.
\\
Case~5.3.2.2.3.2.2.4: $| \lbrace k \in [1, m] : v_k \in V, w_k \in W \rbrace | \geq 3$.
\\
Then, $|V \cup W| \geq 6$, which is impossible.
\\
Case~5.3.2.2.4: $V \subseteq \lbrace w_1, \ldots w_m \rbrace$ and $W \not\subseteq \lbrace v_1, \ldots v_m \rbrace$.
\\
Then, $V |_{\cal D} W = W \cap \lbrace w_1, \ldots, w_m \rbrace = V |_{{\cal D}'} W$.
\\
Case~5.3.2.2.5: $V \subseteq \lbrace w_1, \ldots w_m \rbrace$ and $W \subseteq \lbrace v_1, \ldots v_m \rbrace$.
\\
Similar to Case~5.3.2.2.3.\qed
\end{proof}
We extend the negative results to the ``liberal'' and Hamming properties.
The proof will be based on an adaptation of the Hamster Wheel.
Note that the Hamming distance is a realistic
distance which has been investigated by many researchers.
This strengthen the importance of
Proposition~\ref{DISpascaracham} below in the sense that not only abstract but also concrete cases
do not admit a normal characterization.

\begin{proposition}\label{DISpascaracham}
Let ${\cal L} = \langle {\cal A}, {\cal C} \rangle$ be a propositional language
with ${\cal A}$ infinite and countable,
${\cal M}$ a matrix on $\cal L$,
$\cal V$ the set of all $\cal M$-valuations,
${\cal N}$ the set of all symmetric, HIR, liberally IR, liberally positive, and liberally TIR
distance operators from ${\cal P}({\cal V})^2$ to ${\cal P}({\cal V})$, and
${\cal O}$ a set of distance operators from ${\cal P}({\cal V})^2$ to ${\cal P}({\cal V})$ such that ${\cal N} \subseteq {\cal O}$.
\\
Then, there does not exist a normal characterization of ${\cal O}$.
\end{proposition}

\begin{proof}
Suppose the contrary, i.e. suppose there are $n \in \mathbb{N}^+$ and
a relation $\Phi$ on ${\cal P}({\cal V})^{3n}$ such that
\begin{description}
\item[$(0)$] for every binary operator $|$ on ${\cal P}({\cal V})$,
we have $| \in {\cal O}$ iff
\\
$\forall \: V_1, \ldots, V_n$, $W_1, \ldots, W_n \subseteq {\cal V}$,
\\
$(V_1, \ldots, V_n, W_1, \ldots, W_n, V_1 | W_1, \ldots, V_n | W_n) \in \Phi$.
\end{description}
As ${\cal A}$ is infinite, there are distinct $p_1, \ldots, p_m,$ $q_1, \ldots, q_m$ in ${\cal A}$, with $m = n + 3$.
\\
Let's pose ${\cal M} = \langle T, D, f \rangle$.
\\
As $D \not= \emptyset$ and $T \setminus D \not= \emptyset$, there are distinct $0, 1 \in T$.
\\
Now, $\forall \: i \in [1, m]$, let $v_i$ be the $\cal M$-valuation that assigns $1$ to $p_i$ and $0$ to each other
atom of ${\cal A}$.
\\
Similarly, $\forall \: i \in [1, m]$, let $w_i$ be the $\cal M$-valuation that assigns $1$ to $q_i$ and $0$ to each other
atom of ${\cal A}$.
\\
Let $X = \lbrace v_1, \ldots, v_m, w_1 \ldots, w_m \rbrace$.
\\
Note that $\forall \: v, w \in X$, with $v \not= w$, we have $|h(v, w)| = 2$.
\\
Finally, let ${\cal D}$ be the pseudo-distance on $\cal V$ such that
${\cal D} = \langle \mathbb{R} \cup \lbrace |\mathbb{N}| \rbrace, \prec, d \rangle$,
where $\prec$ and $d$ are defined as follows.
\\
Let $c, c' \in \mathbb{R} \cup \lbrace |\mathbb{N}| \rbrace$.
Then, $c \prec c'$ iff $(c, c' \in \mathbb{R}$ and $c < c')$ or
$(c \in \mathbb{R}$ and $c' = |\mathbb{N}|)$.
\\
Let $v, w \in {\cal V}$ and
consider the cases which follow:
\\
Case~1: $v = w$.
\\
Case~2: $v \not= w$.
\\
Case~2.1: $\lbrace v, w \rbrace \not\subseteq X$.
\\
Case~2.1.1: $|h(v, w)| = 1$.
\\
Case~2.1.2: $|h(v, w)| \geq 2$.
\\
Case~2.2: $\lbrace v, w \rbrace \subseteq X$.
\\
Case~2.2.1: $\lbrace v, w \rbrace \subseteq \lbrace v_1, \ldots, v_m \rbrace$.
\\
Case~2.2.2: $\lbrace v, w \rbrace \subseteq \lbrace w_1, \ldots, w_m \rbrace$.
\\
Case~2.2.3: $\exists \: i, j \in [1, m],\; \lbrace v, w \rbrace = \lbrace v_i, w_j \rbrace$.
\\
Case~2.2.3.1: $i = j$.
\\
Case~2.2.3.2: $abs(i - j) \in \lbrace 1, m-1 \rbrace$.
\\
Case~2.2.3.3: $1 < abs(i - j) < m-1$.
\\
Then,
$$
d(v, w) = \left \{ \begin{array}{l l}
0 & \textrm{if Case~1 holds;}\\
1.4 & \textrm{if Case~2.1.1 holds;}\\
|h(v, w)| & \textrm{if Case~2.1.2 holds;} \\
2.1 & \textrm{if Case~2.2.1 holds;} \\
2.1 & \textrm{if Case~2.2.2 holds;} \\
2.4 & \textrm{if Case~2.2.3.1 holds;}\\
2.5 & \textrm{if Case~2.2.3.2 holds;}\\
2.2 & \textrm{if Case~2.2.3.3 holds.}\\
\end{array} \right .
$$
\\
Note that ${\cal D}$ is an adaptation of the Hamster Wheel of \cite{Schlechta5}.
The reader can find a picture of $\cal D$ in Figure~2.
\\
\begin{center}
\epsfig{file=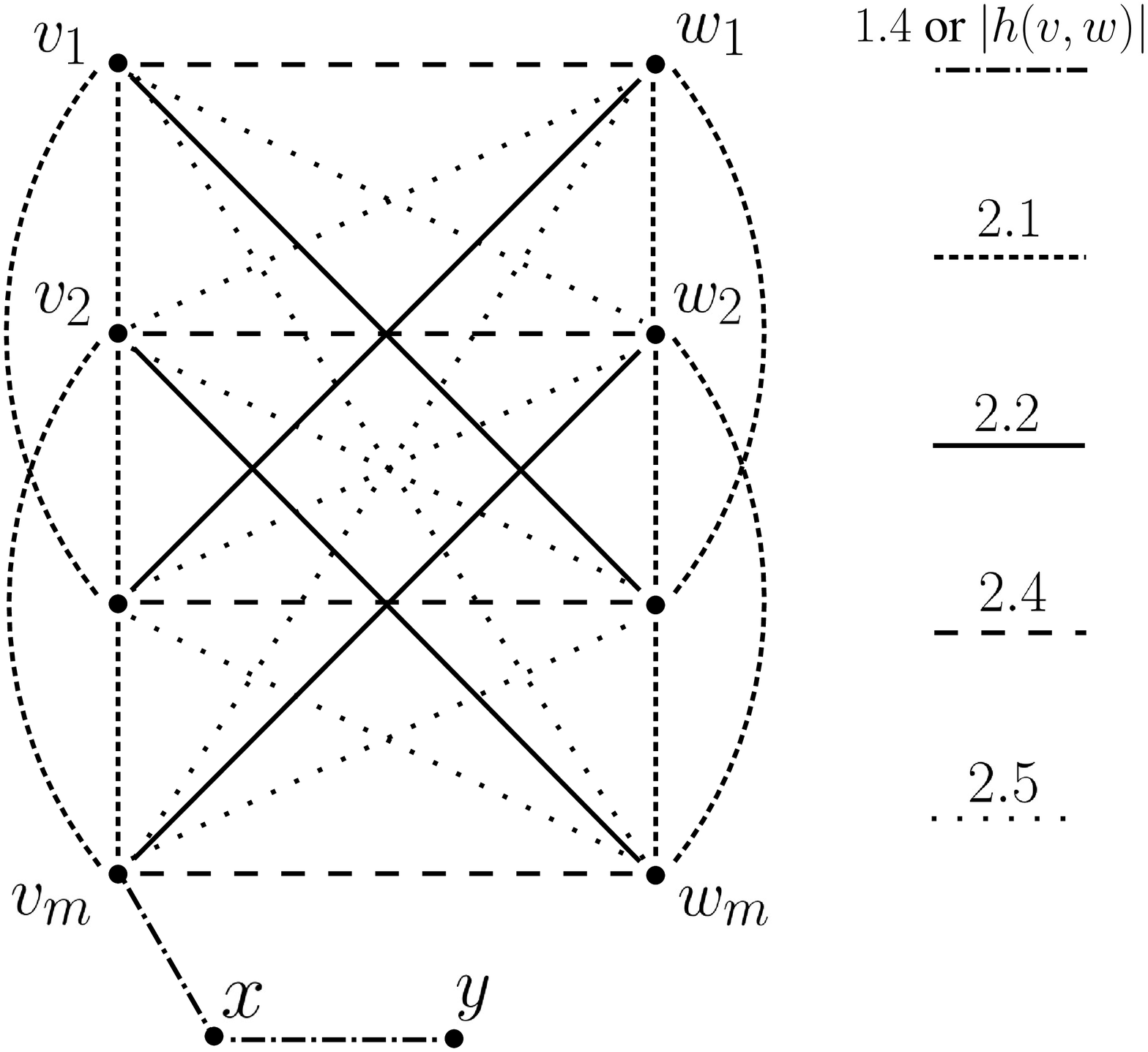, height=8cm}
\hfill \\
Figure 2: An adaptation of Hamster Wheel.
\end{center}
\hfill \\
Let $|$ be the binary operator on ${\cal P}({\cal V})$ defined as follows.
\\
Let $V, W \subseteq {\cal V}$ and
consider the cases that follow:
\\
Case~1: $\forall v \in V$, $\forall w \in W, \lbrace v, w \rbrace \subseteq X$ or $3 \leq |h(v, w)|$.
\\
Case~1.1: $V \cap X = \lbrace v_m, v_1 \rbrace$ and $W \cap X = \lbrace w_m, w_1 \rbrace$.
\\
Case~1.2: $V \cap X = \lbrace w_m, w_1 \rbrace$ and $W \cap X = \lbrace v_m, v_1 \rbrace$.
\\
Case~1.3: $\lbrace V \cap X, W \cap X \rbrace \not= \lbrace \lbrace v_m, v_1 \rbrace,  \lbrace w_m, w_1 \rbrace \rbrace$.
\\
Case~2: $\exists \: v \in V$, $\exists \: w \in W$, $\lbrace v, w \rbrace \not\subseteq X$ and $|h(v, w)| < 3$.
\\
Then,
$$
V | W = \left \{ \begin{array}{l l}
\lbrace w_m \rbrace & \textrm{if Case~1.1 holds;}\\
\lbrace v_m \rbrace & \textrm{if Case~1.2 holds;}\\
V |_{\cal D} W & \textrm{if Case~1.3 or Case~2 holds.}
\end{array} \right .
$$
The difference between $|$ and $|_{\cal D}$ is sufficiently strong so that $|$ is not a distance operator.
The proof is verbatim the same as for $(1)$ in the proof of Proposition~\ref{DISpascarac}.
\\
Consequently, $| \not\in {\cal O}$, thus, by $(0)$, we get that
\begin{description}
\item[$(1)$] $\exists V_1, \ldots, V_n, W_1, \ldots, W_n \subseteq {\cal V}$,
\\
$(V_1, \ldots, V_n, W_1, \ldots, W_n, V_1 | W_1, \ldots, V_n | W_n) \not\in \Phi$.
\end{description}
Moreover, we chose $m$ sufficiently big so that:
\begin{description}
\item[$(2)$] $\exists \: r \in [1, m-1]$,
$\forall \: i \in [1, n]$,
\\
$\lbrace V_i \cap X, W_i \cap X \rbrace \not= \lbrace \lbrace v_r, v_{r+1} \rbrace, \lbrace w_r, w_{r+1} \rbrace \rbrace$.
\end{description}
The proof is verbatim the same as for $(3)$ in the proof of Proposition~\ref{DISpascarac},
except that $V_i$ and $W_i$ are replaced by $V_i \cap X$ and $W_i \cap X$.
\\
Let $|'$ be the binary operator on ${\cal P}({\cal V})$ defined as follows.
\\
Let $V, W \subseteq {\cal V}$ and
consider the cases that follow:
\\
Case~1: $\forall v \in V$, $\forall w \in W$, $\lbrace v, w \rbrace \subseteq X$ or $3 \leq |h(v, w)|$.
\\
Case~1.1: $V \cap X = \lbrace v_r, v_{r+1} \rbrace$ and $W \cap X = \lbrace w_r, w_{r+1} \rbrace$.
\\
Case~1.2: $V \cap X = \lbrace w_r, w_{r+1} \rbrace$ and $W \cap X = \lbrace v_r, v_{r+1} \rbrace$.
\\
Case~1.3: $\lbrace V \cap X, W \cap X \rbrace \not= \lbrace \lbrace v_r, v_{r+1} \rbrace,  \lbrace w_r, w_{r+1} \rbrace \rbrace$.
\\
Case~2: $\exists \: v \in V$, $\exists \: w \in W$, $\lbrace v, w \rbrace \not\subseteq X$ and $|h(v, w)| < 3$.
\\
Then,
$$
V |' W = \left \{ \begin{array}{l l}
\lbrace w_{r+1} \rbrace & \textrm{if Case~1.1 holds;}\\
\lbrace v_{r+1} \rbrace & \textrm{if Case~1.2 holds;}\\
V | W & \textrm{if Case~1.3 or Case~2 holds.}\\
\end{array} \right .
$$
The difference between $|'$ and $|$ is ``invisible'' for $\Phi$.
\\
More formally, $\forall \: i \in [1, n]$, $V_i |' W_i = V_i | W_i$.
\\
The proof is obvious by
$(2)$.
Thus, by $(1)$, we get:
\\
$(V_1, \ldots, V_n$, $W_1, \ldots, W_n$, $V_1 |' W_1, \ldots, V_n |' W_n) \not\in \Phi$.
\\
Therefore, by $(0)$, we get:
\begin{description}
\item[$(3)$] $|' \not\in {\cal O}$.
\end{description}
But, in parallel, there is a convenient
pseudo-distance that represents $|'$. Indeed, let ${\cal D}'$ be the
pseudo-distance on $\cal V$ such that
${\cal D}' = \langle \mathbb{R} \cup \lbrace |\mathbb{N}|\rbrace, \prec, d' \rangle$, where
$d'$ is the function such that $\forall \: v, w \in {\cal V}$,
$$
d'(v, w) = \left \{ \begin{array}{l l}
\!\!\!2.3 & \!\!\!\textrm{if}\: \exists i \in [r+1, m], \lbrace v, w \rbrace = \lbrace v_i, w_i \rbrace;\\
\!\!\!d(v, w) & \!\!\!\textrm{otherwise.}
\end{array} \right .
$$
Note that $\forall \: v, w \in {\cal V}$, we have:
\\
$|h(v,w)| \in \mathbb{N}$ iff $d(v,w) \in \mathbb{R}$ iff $d'(v,w) \in \mathbb{R}$.
\\
Thus, $|h(v,w)| = |\mathbb{N}|$ iff $d(v,w) = |\mathbb{N}|$ iff $d'(v,w) = |\mathbb{N}|$.
\\
Note again that $\forall \: v, w \in {\cal V}$, with $|h(v,w)| \in \mathbb{N}$, we have:
\\
$|h(v, w)| \leq d'(v, w) \leq d(v,w) \leq |h(v, w)| + 0.5$.
\\
We will show:
\begin{description}
\item[$(4)$] $|' = |_{{\cal D}'}$.
\end{description}
But, ${\cal D}'$ is symmetric, liberally IR, and liberally positive.
\\
In addition, we will show:
\begin{description}
\item[$(5)$] ${\cal D}'$ is HIR;
\item[$(6)$] ${\cal D}'$ is liberally TIR.
\end{description}
So, $|'$ is a symmetric, liberally IR, liberally positive, liberally TIR, and HIR
distance operator.
\\
Therefore, $|' \in {\cal N}$ and thus:
\begin{description}
\item[$(7)$] $|' \in {\cal O}$.
\end{description}
Finally, $(3)$ and $(7)$ entail a contradiction.
\\
\\
{\it Proof of $(4)$}. Let $V, W \subseteq {\cal V}$.
\\
Case~1: $\forall v \in V$, $\forall w \in W$, $\lbrace v, w \rbrace \subseteq X$ or $3 \leq |h(v, w)|$.
\\
Case~1.1: $V \cap X = \lbrace v_r, v_{r+1} \rbrace$ and $W \cap X = \lbrace w_r, w_{r+1} \rbrace$.
\\
Then, $V |' W = \lbrace w_{r+1} \rbrace = V |_{{\cal D}'} W$.
\\
Case~1.2: $V \cap X = \lbrace w_r, w_{r+1} \rbrace$ and $W \cap X = \lbrace v_r, v_{r+1} \rbrace$.
\\
Then, $V |' W = \lbrace v_{r+1} \rbrace = V |_{{\cal D}'} W$.
\\
Case 1.3: $V \cap X = \lbrace v_m, v_1 \rbrace$ and $W \cap X = \lbrace w_m, w_1 \rbrace$.
\\
Then, $V |' W = \lbrace w_m \rbrace = V |_{{\cal D}'} W$.
\\
Case 1.4: $V \cap X = \lbrace w_m, w_1 \rbrace$ and $W \cap X = \lbrace v_m, v_1 \rbrace$.
\\
Then, $V |' W = \lbrace v_m \rbrace = V |_{{\cal D}'} W$.
\\
Case~1.5: $\lbrace V \cap X, W \cap X \rbrace \not\in$
\\
$\lbrace \lbrace \lbrace v_m, v_1 \rbrace,  \lbrace w_m, w_1 \rbrace \rbrace$,
$\lbrace \lbrace v_r, v_{r+1} \rbrace,  \lbrace w_r, w_{r+1} \rbrace \rbrace \rbrace$.
\\
Then, $V |' W = V | W = V |_{\cal D} W$.
\\
Case~1.5.1: $V \cap W \not= \emptyset$.
\\
Then, $V |_{\cal D} W = V \cap W = V |_{{\cal D}'} W$.
\\
Case~1.5.2: $V \cap W = \emptyset$.
\\
Case~1.5.2.1: $V \cap X = \emptyset$ or $W \cap X = \emptyset$.
\\
Then,
$\forall \: v \in V$, $\forall \: w \in W$, $d'(v, w) = d(v, w)$.
\\
Therefore, $V |_{\cal D} W = V |_{{\cal D}'} W$.
\\
Case~1.5.2.2: $V \cap X \not= \emptyset$ and $W \cap X \not= \emptyset$.
\\
Then, we will show:
\begin{description}
\item[$(4.1)$] $V |_{\cal D} W = V \cap X |_{\cal D} W \cap X$;
\item[$(4.2)$] $V |_{{\cal D}'} W = V \cap X |_{{\cal D}'} W \cap X$.
\end{description}
But, we have $V \cap X |_{\cal D} W \cap X = V \cap X |_{{\cal D}'} W \cap X$.
\\
The proof of this is verbatim the same as for Case~5.3.2.2,
in the proof of $(5)$, in the
proof of Proposition~\ref{DISpascarac}, except that
$V$ and $W$ are replaced by $V \cap X$ and $W \cap X$.
\\
Case~2: $\exists \: v \in V$, $\exists \: w \in W$, $\lbrace v, w \rbrace \not\subseteq X$ and $|h(v, w)| \leq 2$.
\\
Then, $V |' W = V | W = V |_{\cal D} W$.
\\
Case~2.1. $V \cap W \not= \emptyset$.
\\
Then, $V |_{\cal D} W = V \cap W = V |_{{\cal D}'} W$.
\\
Case~2.2. $V \cap W = \emptyset$.
\\
Case~2.2.1. $\exists \: v' \in V$, $\exists \: w' \in W$, $|h(v, w)| = 1$.
\\
Then, $V |_{\cal D} W = \lbrace w \in W : \exists \: v \in V, |h(v, w)| = 1 \rbrace = V |_{{\cal D}'} W$.
\\
Case~2.2.2. $\forall \: v' \in V$, $\forall \: w' \in W$, $|h(v, w)| \geq 2$.
\\
Then, $V |_{\cal D} W =$
\\
$\lbrace w \in W : \exists \: v \in V, \lbrace v, w \rbrace \not\subseteq X$ and
$|h(v, w)| = 2 \rbrace = V |_{{\cal D}'} W$.\hfill
\\
\\
{\it Proof of $(4.1)$}. Direction: ``$\subseteq$''.
\\
Let $w \in V |_{\cal D} W$.
\\
Then, $\exists \: v \in V$, $\forall \: v' \in V$, $\forall \: w' \in W$, $d(v,w) \preceq d(v', w')$.
\\
Case~1: $\lbrace v, w \rbrace \subseteq X$.
\\
Then, $w \in V \cap X |_{\cal D} W \cap X$.
\\
Case~2: $\lbrace v, w \rbrace \not\subseteq X$.
\\
We have $\exists \: v' \in V \cap X$ and $\exists \: w' \in W \cap X$.
\\
In addition, $d(v', w') \in \mathbb{R}$ and $d(v', w') \leq 2.5$.
\\
Case~2.1: $|h(v,w)| = |\mathbb{N}|$.
\\
Then, $d(v,w) = |\mathbb{N}|$.
\\
Therefore, $d(v', w') \prec d(v, w)$, which is impossible.
\\
Case~2.2: $|h(v,w)| \in \mathbb{N}$.
\\
Then, $d(v, w) \in \mathbb{R}$ and $3 \leq |h(v, w)| \leq d(v,w)$.
\\
Therefore, $d(v', w') < d(v, w)$.
\\
Thus, $d(v', w') \prec d(v, w)$, which is impossible.

Direction: ``$\supseteq$''.
\\
Let $w \in V \cap X |_{\cal D} W \cap X$.
\\
Then, $\exists \: v \in V \cap X$ such that
\\
$\forall \: v' \in V \cap X$, $\forall \: w' \in W \cap X$, $d(v, w) \preceq d(v', w')$.
\\
Let $v' \in V$, $w' \in W$.
\\
Case~1: $\lbrace v', w' \rbrace \subseteq X$.
\\
Then, $d(v, w)  \preceq d(v', w')$.
\\
Case~2: $\lbrace v', w' \rbrace \not\subseteq X$.
\\
As $v, w \in X$, we have
$d(v, w) \in \mathbb{R}$ and $d(v, w) \leq 2.5$.
\\
Case~2.1: $|h(v', w')| = |\mathbb{N}|$.
\\
Then, $d(v', w') = |\mathbb{N}|$. Thus,
$d(v, w) \prec d(v', w')$.
\\
Case~2.2: $|h(v', w')| \in \mathbb{N}$.
\\
Then,
$d(v', w') \in \mathbb{R}$ and $3 \leq |h(v', w')| \leq d(v', w')$.
\\
Therefore, $d(v,w) < d(v', w')$.
\\
Thus, $d(v,w) \prec d(v', w')$.
\\
Consequently, in all cases, $d(v, w) \preceq d(v', w')$.
\\
Thus, $w \in V |_{\cal D} W$.
\\
\\
{\it Proof of $(4.2)$}. Verbatim the proof of $(4.1)$, except that
$|_{\cal D}$ and $d$ are replaced by $|_{{\cal D}'}$ and $d'$.
\\
\\
{\it Proof of $(5)$}.
Let $v, w, x \in {\cal V}$ with $|h(v, w)| < |h(v, x)|$.
\\
Case~1: $|h(v, x)| = |\mathbb{N}|$.
\\
Then, $|h(v,w)| \in \mathbb{N}$.
\\
Thus, $d'(v, w) \in \mathbb{R}$ and $d'(v, x) = |\mathbb{N}|$.
\\
Therefore, $d'(v, w) \prec d'(v, x)$.
\\
Case~2: $|h(v,x)| \in \mathbb{N}$.
\\
Then, $|h(v,w)| \in \mathbb{N}$.
\\
Therefore, $d'(v,x) \in \mathbb{R}$, $d'(v,w) \in \mathbb{R}$, and
$d'(v, w) \leq |h(v, w)| + 0.5 < |h(v, w)| + 1 \leq |h(v, x)| \leq d'(v, x)$.
\\
Thus, $d'(v, w) \prec d'(v, x)$.
\\
\\
{\it Proof of $(6)$}. Let $v, w, x \in {\cal V}$.
\\
Note that $h(v,x) \subseteq h(v, w) \cup h(w,x)$.
\\
Therefore, $|h(v,x)| \leq |h(v, w) \cup h(w,x)|$.
\\
Case~1: $d'(v,x) = |\mathbb{N}|$.
\\
Then, $|h(v,x)| = |\mathbb{N}|$.
\\
Now, suppose $d'(v, w) \in \mathbb{R}$ and $d'(w, x) \in \mathbb{R}$.
\\
Then, $|h(v,w)|, |h(w,x)| \in \mathbb{N}$.
\\
Thus,
$|h(v, w) \cup h(w, x)| \in \mathbb{N}$.
\\
Therefore, $|h(v,x)| \in \mathbb{N}$, which is impossible.
\\
Thus, $d'(v, w) = |\mathbb{N}|$ or $d'(w, x) = |\mathbb{N}|$.
\\
Case~2: $d'(v,x), d'(v,w), d'(w,x) \in \mathbb{R}$.
\\
Case~2.1: $|h(v, w)| = 0$ or $|h(w, x)| = 0$. Trivial.
\\
Case~2.2: $|h(v, w)| \geq 1$ and $|h(w, x)| \geq 1$.
\\
Case~2.2.1: $|h(v, w)| \geq 2$ or $|h(w, x)| \geq 2$.
\\
Case~2.2.1.1: $|h(v, x)| \in \lbrace 0, 1, 2 \rbrace$.
\\
Then, $d'(v, x) \leq |h(v, x)| + 0.5 \leq 2.5 < 3 \leq
|h(v, w)| + |h(w, x)| \leq d'(v, w) + d'(w, x)$.
\\
Case~2.2.1.2: $|h(v, x)| \geq 3$.
\\
Then, $d'(v, x) = |h(v, x)| \leq |h(v, w)| + |h(w, x)| \leq d'(v, w) + d'(w, x)$.
\\
Case~2.2.2: $|h(v, w)| = 1$ and $|h(w, x)| = 1$.
\\
Case~2.2.2.1: $|h(v, x)| \in \lbrace 0, 1, 2 \rbrace$.
\\
Then, $d'(v, x) \leq |h(v, x)| + 0.5 \leq 2.5 < 1.4 + 1.4 
= d'(v, w) + d'(w, x)$.
\\
Case~2.2.2.2: $|h(v, x)| \geq 3$.
\\
Then, $|h(v, x)| > |h(v, w)| + |h(w, x)|$, impossible.\qed
\end{proof}

\section{Conclusion} \label{DISconclu}

We laid the focus on the question to know whether $(\star loop)$ can be replaced
by a finite condition in Proposition~\ref{DIScaracrevclass}.
Obviously, the presence of $(\star loop)$ is due to the presence of $(| loop)$.
So, to solve the problem one might attack its source, i.e. try to replace $(| loop)$ by a finite condition
in Proposition~\ref{DISalgebraic}.
But, we showed in the present paper that for families of distance operators,
there is no normal characterization.
The symmetric family is concerned with this and therefore $(| loop)$ cannot be
replaced by a finite and universally quantified condition.

Now, we can go further. Indeed, there is a strong connexion between the distance operators
and the distance-based revision operators.
Lehmann {\it et al.} used this connexion to get their results on the latter from their results on the former. It is reasonable to think that the same thing can be done with our negative results, i.e
this paper can certainly be continued in future work to show that for families
of distance-based revision operators, there is no normal characterization either.
For instance, the family which is symmetric, CP, and DP might well be concerned with this,
which suggests that $(\star loop)$ cannot be replaced by a finite and universally quantified condition.

In addition, this direction for future work can still be followed if
we define the distance-based revision in a non-classical framework. Indeed, as Lehmann {\it et al.} did,
we worked in a general framework. For instance, if we define the revision
in the $\cal FOUR$ framework ---$\cal FOUR$ is a well-known paraconsistent logic from \cite{Belnap1} and \cite{Belnap2} --- then we can probably use the results of \cite{LehmannMagidorSchlechta1}
and our results respectively to
show characterizations of revision operators and show that they cannot be really improved.

Moreover, most of the approaches to belief revision treat in a trivial way
inconsistent sets of beliefs (if they are treated at all).
However, people may be rational despite inconsistent beliefs
(there may be overwhelming evidence for both something and its contrary). There
are also inconsistencies in principle impossible
to eliminate like the ``Paradox of the Preface'' \cite{Makinson3}.
The latter says that a conscientious author has reasons to believe that everything
written in his book is true. But, because of human imperfection,
he is sure that his book contains errors, and thus that something must be false.
Consequently, he has (in the absolute sense) both reasons to believe
that everything is true and that something is false.
So, principles of rational belief revision must work on inconsistent sets of beliefs.
Standard approaches to belief revision (e.g. AGM) all fail to do this as they are based on classical logic.
Paraconsistent logics (like e.g. $\cal FOUR$) could be the bases of more adequate approaches.

Another advantage of such approaches is that they will not be forced
to eliminate a contradiction even when there is no good way to do it.
Contradictions could be tolerated until new information eventually comes to justify
one or another way of elimination.

Finally, such approaches will benefit from an extended field of application
which includes multi-agent systems where the agents can have individually
inconsistent beliefs. Furthermore, it is easy to see that these perspectives for belief revision
can be transposed to belief merging.

\bibliographystyle{aaai}
\bibliography{generalBennaim}

\end{document}